\documentclass[aps,preprint,twocolumn,tightenlines,showpacs,amsmath,amssymb,nofootinbib,floatfix,superscriptaddress]{revtex4}

\usepackage{graphicx}
\usepackage{amsfonts}
\usepackage{color}

\begin{document}

\title{The low lying scalar resonances in the $D^0$ decays into $K^0_s$ and $f_0(500)$, $f_0(980)$, $a_0(980)$}

\date{\today}

\author{Ju-Jun Xie}
\affiliation{Institute of Modern Physics, Chinese Academy of
Sciences, Lanzhou 730000, China} \affiliation{State Key Laboratory
of Theoretical Physics, Institute of Theoretical Physics, Chinese
Academy of Sciences, Beijing 100190, China}

\author{L.R. Dai}
\affiliation{Department of Physics, Liaoning Normal University,
Dalian 116029, China}

\author{E.~Oset}
\affiliation{Institute of Modern Physics, Chinese Academy of
Sciences, Lanzhou 730000, China} \affiliation{Departamento de
F\'{\i}sica Te\'orica and IFIC, Centro Mixto Universidad de
Valencia-CSIC Institutos de Investigaci\'on de Paterna, Aptdo.
22085, 46071 Valencia, Spain}

\begin{abstract}

The $D^0$ decay into $K^0_s$ and a scalar resonance, $f_0(500)$,
$f_0(980)$, $a_0(980)$, is studied obtaining the scalar resonances
from final state interaction of a pair of mesons produced in a first
step in the $D^0$ decay into $K^0_s$ and the pair of pseudoscalar
mesons. This weak decay is very appropriate for this kind of study
because it allows to produce the three resonances in the same decay
in a process that is Cabibbo allowed, hence the rates obtained are
large compared to those of $\bar{B}^0$ decays into $J/\psi$ and a
scalar meson that have at least one Cabibbo suppressed vertex.
Concretely the $a_0(980)$ production is Cabibbo allowed here, while
it cannot be seen in the $\bar{B}^0_s$ decay into $J/\psi a_0(980)$
and is doubly Cabibbo suppressed in the $\bar{B}^0$ decay into
$J/\psi a_0(980)$ and has not been identified there. The fact that
the three resonances can be seen in the same reaction, because there
is no isospin conservation in the weak decays, offers a unique
opportunity to test the ideas of the chiral unitary approach where
these resonances are produced from the interaction of pairs of
pseudoscalar mesons.

\end{abstract}


\maketitle

\section{Introduction}

The rates for $D^0$ decay into $K^0_s$ and a scalar resonance,
$f_0(980)$, $a_0(980)$ are measured by the CLEO collaboration in
Ref.~\cite{Muramatsu:2002jp} and Ref.~\cite{Rubin:2004cq}
respectively and the rates are relatively large. The  $f_0(980)$ is
seen through its decay into $\pi^+ \pi^-$ and the $a_0(980)$ through
the $\pi^0 \eta$ channel. Related references on the issue can be
seen in the PDG \cite{pdg}. Theoretical work on these decays is
scarce and is mostly devoted to issues related to CP violation or
$D^0-D^{*0}$ mixing. In Ref.~\cite{kaminski} a thorough study is
done of the $D^0 \to K^0_s \pi^+ \pi^-$ reaction and the amplitude
is parametrized in terms of form factors, resonance parameters and
different couplings, amounting to a set of 33 free parameters, which
are fitted to the Belle~\cite{Poluektov:2010wz} and
BaBar~\cite{delAmoSanchez:2010xz} data. The purpose is to have a
good amplitude that can be used to determine the $D^0-D^{*0}$ mixing
parameters and the Cabibbo-Kobayashi-Maskawa (CKM) angle $\gamma$.

The aim of the present work is different, we only evaluate the part
of the $D^0 \to K^0_s \pi^+ \pi^-$ amplitude corresponding to a
$K^0_s$ and two pions propagating in $s$-wave, which will show the
$f_0(500)$ and $f_0(980)$ resonances. In addition we study the $D^0
\to K^0_s \pi^0 \eta$ amplitude, where the $a_0(980)$ resonance
shows up, and relate it to the former one. However, we show that, by
using basic symmetries and the chiral unitary approach to deal with
the meson meson interaction in coupled channels, one is able to
determine the shapes of the different amplitudes and the relative
weight to each other with no free parameters. Hence genuine
predictions for the shapes of these amplitudes and the relative
weights of  $f_0(500)$, $f_0(980)$ and $a_0(980)$ can be made and
compared with experiment.

The chiral unitary approach for meson meson interaction makes use of
the Bethe Salpeter (BS) equation in coupled channels. One takes all
possible meson meson channels that couple within $SU(3)$ to certain
given quantum numbers and the BS equation guaranties exact unitary.
The kernel (potential) for the BS equation is taken from the chiral
Lagrangians \cite{Gasser:1983yg,Bernard:1995dp} and there is freedom
for only some regularization scale in the meson meson loops, which
is fitted to the meson meson scattering data. A good agreement with
experimental data is obtained up to 1.2 GeV
\cite{npa,ramonet,kaiser,markushin,juanito,rios}. One of the
consequences of this approach is that the resonances  $f_0(500)$,
$f_0(980)$, $a_0(980)$ and $\kappa(800)$ are automatically generated
from these potentials and the use of the BS equations. In this way
these resonances qualify as dynamically generated states, some kind
of composite, or molecular, meson meson states, in the same way as
the deuteron qualifies as a bound state of a proton and a neutron
and not a more exotic object~\cite{Weinberg:1965zz}.  The approach
not only provides the meson meson amplitudes but has been tested
successfully in virtually any reaction where any of the former
resonances is produced. The latest test was the study of the $B^0$
and $B^0_s$ decays into $J/\psi f_0(500)$ and $J/\psi f_0(980)$
which was done in Ref.~\cite{weihong} (a list of different reactions
where the former resonances are produced can also be found there),
where a natural explanation was given of the observed facts that the
$\bar{B}^0_s$ decays into $J/\psi f_0(980)$, while no signal is seen
for $J/\psi f_0(500)$, and the $\bar{B}^0$ decays into $J/\psi
f_0(500)$ and only a small fraction is seen for the $J/\psi
f_0(980)$.

The $D^0$ decay into $K^0_s$ and a scalar resonance, $f_0(500)$,
$f_0(980)$, $a_0(980)$ is a privileged case to test the nature of
these resonances. Indeed, as we shall see, the three processes are
Cabibbo allowed and the rates of production are big compared to
those of the $\bar{B}^0$ decays into $J/\psi$ and one of these
resonances, where necessarily one of the vertices, the $V_{cb}$, is
Cabibbo suppressed \cite{Aaij:2011fx,Aaij:2013zpt,Stone:2013eaa}. On
the other hand, the $a_0(980)$ has not been reported in $\bar{B}^0$,
$\bar{B}^0_s$ decays. As one can see in
Ref.~\cite{Stone:2013eaa,weihong}, in the decay of $\bar{B}^0_s$
into $J/\psi$ one gets an extra $s \bar s$ pair that has $I=0$ and
does not allow the $a_0(980)$ production upon hadronization. On the
other hand in the $B^0$ decay into $J/\psi$ one gets an extra $d
\bar d$ pair that could lead to the $a_0(980)$ upon hadronization,
but the process is doubly Cabibbo suppressed. It is found there that
a signal is seen for the $f_0(500)$ production and only a small
fraction is reported for $f_0(980)$ production~\cite{Aaij:2013zpt}.
One should expect also a minor rate for $a_0(980)$ production in
this case and, in fact, this mode of decay is not reported.  In the
present case the $a_0(980)$ production is allowed and the rates are
large~\cite{Rubin:2004cq}. The fact that we have now weak
interactions that allow for isospin violation permit that both the
$f_0(980)$ and $a_0(980)$ resonances are produced in the same
reaction. This is a novelty with respect to strong interactions that
are isospin conserving. The present weak decay presents then a new
challenge since one can determine the relative weight of production
of each one of these resonances in the same reaction, a new
situation with respect to what one has in strong interaction
reactions.

\section{Formalism}

\begin{figure}[htbp]
\begin{center}
\includegraphics[scale=0.8]{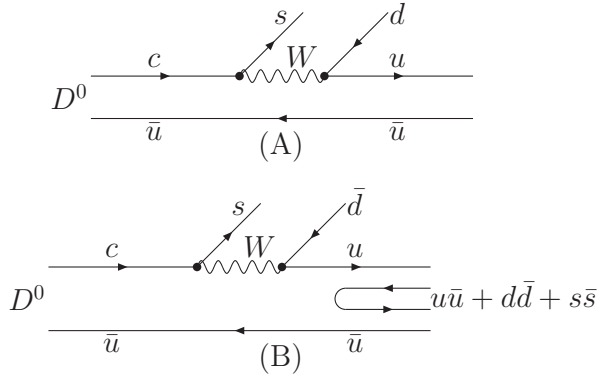}
\caption{(A): Dominant diagrams for $D^0 \to \bar{K}^0 u \bar{u})$
and (B): hadronization of the $u \bar{u}$ to give two mesons.}
\label{Fig:feyn}
\end{center}
\end{figure}

The process for $D^0 \to K^0_s R$ proceeds at the elementary quark
level as depicted in Fig.~\ref{Fig:feyn} (A). The process is Cabibbo
allowed, the $s\bar{d}$ pair produces the $\bar{K}^0$, which will
convert to the observed $K^0_s$ through time evolution with the weak
interaction. The remaining $u\bar{u}$ pair gets hadronized adding an
extra $\bar{q}q$ with the quantum mumbers of the vacuum, $\bar{u}u +
\bar{d}d + \bar{s}s$. This topology is the same as for the
$\bar{B}_s \to J/\psi s\bar{s}$ (substituting the $s\bar{d}$ by
$c\bar{c}$)~\cite{Stone:2013eaa}, that upon hadronization of the
$s\bar{s}$ pair leads to the production of the
$f_0(980)$~\cite{weihong}, which couples mostly to the hadronized $K
\bar{K}$ components.

The hadronization is implemented in an easy way following the work
of Ref.~\cite{alberzou}. One starts with the $q\bar{q}$ matrix $M$

\begin{equation}\label{eq:1}
M=\left(
           \begin{array}{ccc}
             u\bar u & u \bar d & u\bar s \\
             d\bar u & d\bar d & d\bar s \\
             s\bar u & s\bar d & s\bar s \\
           \end{array}
         \right)
\end{equation}
which has the property
\begin{equation}\label{eq:2}
M\cdot M=M \times (\bar{u}u + \bar{d}d + \bar{s}s).
\end{equation}

Hence the $u\bar{u}$ component of Fig.~\ref{Fig:feyn} (B) can be
written as,
\begin{equation}
u\bar{u}(\bar{u}u + \bar{d}d + \bar{s}s) = (M \cdot M)_{11}.
\end{equation}

Next, we rewrite the $q\bar{q}$ matrix $M$ in terms of meson
components, and we have $M$ corresponding to the matrix
$\phi$~\cite{bramon,palomar,gamermann}
\begin{widetext}
\begin{equation}\label{eq:phimatrix}
\phi = \left(
           \begin{array}{ccc}
             \frac{1}{\sqrt{2}}\pi^0 + \frac{1}{\sqrt{3}}\eta + \frac{1}{\sqrt{6}}\eta' & \pi^+ & K^+ \\
             \pi^- & -\frac{1}{\sqrt{2}}\pi^0 + \frac{1}{\sqrt{3}}\eta + \frac{1}{\sqrt{6}}\eta' & K^0 \\
            K^- & \bar{K}^0 & -\frac{1}{\sqrt{3}}\eta + \sqrt{\frac{2}{3}}\eta' \\
           \end{array}
         \right)
\end{equation}
\end{widetext}
This matrix corresponds to the ordinary one used in chiral
perturbation theory~\cite{Gasser:1983yg} with the addition of
$\frac{1}{\sqrt{3}}diag(\eta_1,\eta_1,\eta_1)$ where $\eta_1$ is a
singlet of $SU(3)$, taking into account the standard mixing between
$\eta$ and $\eta'$. The term
$\frac{1}{\sqrt{3}}diag(\eta_1,\eta_1,\eta_1)$ is omitted in the
chiral Lagrangians because the $[\phi,\partial_{\mu}\phi]$ structure
of the Lagrangians renders this term inoperative. In
Ref.~\cite{weihong} the ordinary $\phi$ matrix of chiral
perturbation theory was also used. Here we consider the full $\phi$
matrix of Eq.~(\ref{eq:phimatrix}) since we are concerned with
physical $\eta$ plus $\pi^0$ production.

Hence upon hadronization of the $u\bar{u}$ component we shall have
\begin{eqnarray}
&& u\bar{u}(\bar{u}u + \bar{d}d + \bar{s}s)  \equiv  (\phi \cdot
\phi)_{11} = \frac{1}{2}\pi^0 \pi^0 \nonumber \\
&& + \frac{1}{3} \eta \eta + \frac{2}{\sqrt{6}} \pi^0 \eta  +
\pi^+\pi^- + K^+ K^- , \label{eq:phiphi11}
\end{eqnarray}
where we have omitted the $\eta'$ term because of its large mass.
This means that upon hadronization of the $u\bar{u}$ component we
have $D^0 \to \bar{K}^0 PP$, where $PP$ are the different meson
meson components of Eq.~(\ref{eq:phiphi11}). This is only the first
step, because now these mesons will interact among themselves
delivering the desired meson pair component at the end: $\pi^+
\pi^-$ for the case of the $f_0(500)$ and $f_0(980)$, and $\pi^0
\eta$ for the case of the $a_0(980)$.

\begin{figure*}[htbp]
\begin{center}
\includegraphics[scale=1.]{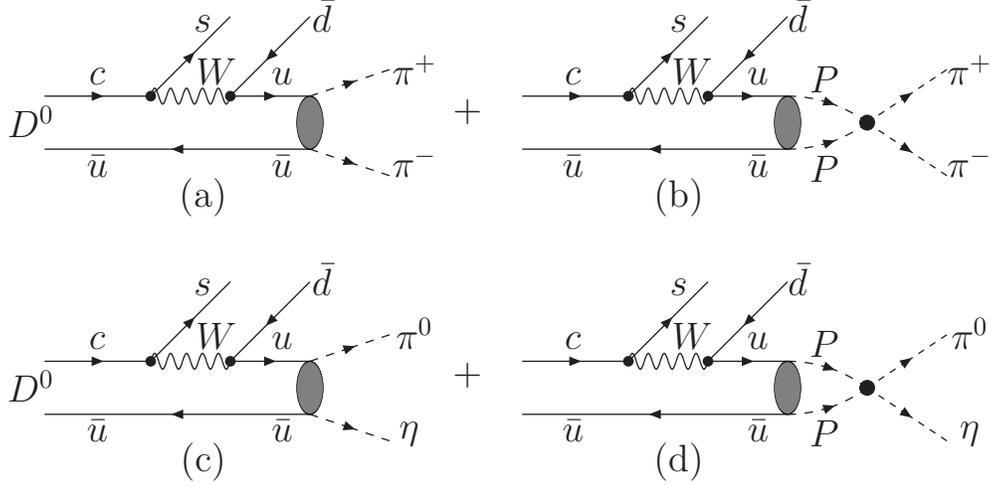}
\caption{ Diagrammatic representation of $\pi^+ \pi^-$ and $\pi^0
\eta$ production. (a) direct $\pi^+ \pi^-$ production, (b) $\pi^+
\pi^-$ production through primary production of a $PP$ pair and
rescattering, (c) primary $\pi^0 \eta$ production, (d) $\pi^0 \eta$
produced through rescattering.} \label{Fig:mesonmesonFSI}
\end{center}
\end{figure*}

The multiple scattering of the mesons is readily taken into account
as shown diagrammatically in Fig.~\ref{Fig:mesonmesonFSI}.

Analytically we shall have
\begin{eqnarray}
&& t(D^0 \to \bar{K}^0 \pi^+ \pi^-) = V_P (1 + G_{\pi^+
\pi^-}t_{\pi^+\pi^- \to \pi^+\pi^-} \nonumber \\
&& + \frac{1}{2} \frac{1}{2} G_{\pi^0\pi^0} t_{\pi^0 \pi^0 \to \pi^+
\pi^-} + \frac{1}{3} \frac{1}{2}G_{\eta \eta} t_{\eta \eta \to \pi^+
\pi^-} \nonumber \\
&& + G_{K^+K^-} t_{K^+ K^- \to \pi^+ \pi^-}), \label{eq:fzero}
\end{eqnarray}
and
\begin{eqnarray}
 && \!\!\!\!\!\!\!\!\! t(D^0 \to \bar{K}^0 \pi^0 \eta) = V_P
(\sqrt{\frac{2}{3}} + \nonumber \\ && \!\!\!\!\!\!\!\!\!
\sqrt{\frac{2}{3}} G_{\pi^0 \eta} t_{\pi^0 \eta \to \pi^0 \eta}  +
G_{K^+K^-} t_{K^+ K^- \to \pi^0 \eta}), \label{eq:azero}
\end{eqnarray}
where $V_P$ is a production vertex, containing the dynamics which is
common to all the terms. $G$ is the loop function of two
mesons~\cite{npa} and $t_{ij}$ are the transition scattering
matrices between pairs of pseudoscalars~\cite{npa}. The $f_0(500)$,
$f_0(980)$, and $a_0(980)$ are produced in $s$-wave where $\pi^0
\pi^0$, $\pi^+ \pi^-$ have isospin $I=0$, hence these terms do not
contribute to $\pi^0 \eta$ production ($I=1$) in
Eq.~(\ref{eq:azero}). Note that in Eq.~(\ref{eq:fzero}) we introduce
the factor $\frac{1}{2}$ extra for the identity of the particles for
$\pi^0 \pi^0$ and $\eta \eta$.

The $t$ matrix is obtained as
\begin{equation}\label{eq:BSeq}
t = [1-VG]^{-1} V,
\end{equation}
where $V_{ij}$ are the transition potentials evaluated in
Refs.~\cite{npa,danydan}. Explicit expressions for $I=0$ are given
in Ref.~\cite{weihong}. We have the $I=1$ case new here and we
present the matrix elements below
\begin{eqnarray}
&& \!\!\! \! \!\! \! \! V_{K^+ K^- \to \pi^0 \eta} \!  = \frac{-
\sqrt{3}}{12f^2} \! ( 3s \!\! - \! \! \frac{8}{3}m^2_K \!\! - \!\! \frac{1}{3}m^2_{\pi} \!\! - \!\! m^2_{\eta} ), \\
&& \!\!\! \! \!\! \! \! V_{K^0\bar{K}^0 \to \pi^0 \eta} = - V_{K^+ K^- \to \pi^0 \eta} ,\\
&& \!\!\! \! \!\! \! \! V_{\pi^0 \eta \to \pi^0 \eta} = -\frac{1}{3f^2}m^2_{\pi}, \\
&& \!\!\! \! \!\! \! \! V_{K^+K^- \to K^+K^-} = -\frac{1}{2f^2} s , \\
&& \!\!\! \! \!\! \! \! V_{K^+K^- \to K^0 \bar{K}^0} = -\frac{1}{4f^2} s , \\
&& \!\!\! \! \!\! \! \! V_{K^0 \bar{K}^0 \to K^0 \bar{K}^0} =
-\frac{1}{2f^2} s ,
\end{eqnarray}
with $f$ the pion decay constant, $f = 93$ MeV, and $s$ is invariant
mass square of the meson-meson system.

The loop function $G$~\cite{npa} is regularized by means of a cut
off. When the $\eta \eta$ channel is explicitly taken into account
the cut off needed is smaller than in Ref.~\cite{npa} and we
follow~\cite{weihong} where it was taken equal to $q_{\rm max} =
600$ MeV.

Finally, the mass distribution for the decay is given
by~\footnote{The decay amplitude $t_{\bar D^0 \to \bar K^0 \pi^+
\pi^-}$ depends on the invariant mass, $M_{\rm inv} = \sqrt{s}$, of
the meson-meson system.}
\begin{equation}\label{eq:dGamma}
\frac{d \Gamma}{d M_{\rm inv}}=\frac{1}{(2\pi)^3}\frac{ p_{\bar K^0}
\tilde{p}_{\pi} }{4M_{D^0}^2} \left| t_{ D^0 \to \bar K^0 \pi^+
\pi^-} \right|^2,
\end{equation}
where $p_{\bar K^0}$ is the $\bar{K}^0$ momentum in the global CM
frame ($D^0$ at rest) and $\tilde{p}_{\pi}$ is the pion momentum in
the $\pi^+ \pi^-$ rest frame,
\begin{eqnarray}\label{eq:pJpsi}
p_{\bar{K}^0} &=& \frac{\lambda^{1/2}(M_{D^0}^2, M_{\bar K^0}^2,
M_{\rm inv}^2)}{2M_{D^0}}, \\
\tilde{p}_{\pi} &=& \frac{\lambda^{1/2}(M_{\rm inv}^2, m_{\pi}^2,
m_{\pi}^2)}{2M_{\rm inv}}  ,
\end{eqnarray}
and similarly for the case of the $\pi^0 \eta$ production.

Before closing this section we should mention that in a three hadron
final state one must look for the interaction of three particles,
for which one must in principle deal with Faddeev
equations~\cite{Faddeev:1960su}. Most of the applications of Faddeev
equations are done for three baryon systems but calculations for
three mesons are becoming available~\cite{albertokan}. However, for
the purpose of the present work it is instructive to follow the idea
in Ref.~\cite{dosreis} for the analogous $D^+ \to K^- \pi^+ \pi^+$
reaction. In this work two body unitarity is imposed on the two body
systems and diagrams related to three body unitarity are evaluated
perturbatively. They are found relevant close to threshold but fade
away rapidly of higher energies. What we have done is in this line
and we have unitarized the $\pi^+\pi^-$, $\pi^0 \eta$ (and coupled
channels pairs) but the $\bar{K}^0$ has been left as a spectator. In
principle we should also look at the interaction of $\bar{K}^0
\pi^-$ which can lead to the $\kappa$ resonance~\cite{ramonet}, yet
the topology of Fig.~\ref{Fig:mesonmesonFSI} (a) does not favor
$s$-wave interaction of $\bar{K}^0 \pi^-$. And furthermore, the
$\kappa$ can also come from a different topology of the diagrams
than those considered in Fig.~\ref{Fig:mesonmesonFSI} (a) for
instance producing a $\pi^+$ meson from the $c$ quark via direct
conversions of $W$ into $\pi^+$ (see section IV,
Fig.~\ref{wexchange} (A)). This is why the $\kappa$ is better seen
in the $D^+ \to K^- \pi^+ \pi^+$ reaction, as discussed in
Ref.~\cite{dosreis}. We do not consider the $\pi K$ interaction
leading to the $\kappa$, with the argument that the $\kappa$, being
a very broad resonance in the $\pi K$ invariant mass, only
contributes a smooth background below the $\pi^+ \pi^-$, or $\pi^0
\eta$ invariant mass distribution when one looks for the $f_0(980)$
or $a_0(980)$ signals and is taken into account in experimental
analysis of these two latter resonances. In this sense, the diagram
of Fig.~\ref{Fig:feyn} chosen and the interaction that we have
considered is also what corresponds to the $K^0_s [\pi^+ \pi^-]_s$,
$M_2$ amplitude of Ref.~\cite{kaminski}, the one that considers the
$s$-wave interaction of the pions and the $f_0(500)$ and $f_0(980)$
resonances, or the $a_0(980)$ when we consider in addition the
$K^0_s [\pi^0 \eta]_s$ amplitude.

\section{Results}

\begin{figure}[htbp]
\begin{center}
\includegraphics[scale=0.44]{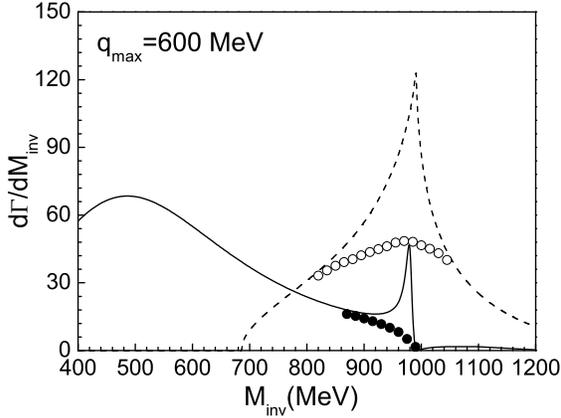}
\caption{The $\pi^+ \pi^-$ (solid line) and $\pi^0 \eta$ (dashed
line) invariant mass distributions for the $D^0 \to \bar{K}^0 \pi^+
\pi^-$ decay and $D^0 \to \bar{K}^0 \pi^0 \eta$ decay, respectively.
A smooth background is plotted below the $a_0(980)$ and $f_0(980)$
peaks.} \label{Fig:dgamr600}
\end{center}
\end{figure}

In Fig.~\ref{Fig:dgamr600}, we show the results of our calculation.
We have taken the cut off $q_{\rm qmax} = 600$ MeV as in
Ref.~\cite{weihong}. We superpose the two mass distributions
$d\Gamma /dM_{\rm inv}$ for $\pi^+ \pi^-$ (solid line) and $\pi^0
\eta$ (dashed line). The scale is arbitrary, since it corresponds to
taking $V_p = 1000$ in Eqs.~(\ref{eq:fzero}) and (\ref{eq:azero}),
but it is the same for the two distributions, which allows us to
compare $f_0(980)$ with $a_0(980)$ production. As we discussed
before, it is a benefit of the weak interactions that we can see
simultaneously both the $I=0$ $f_0(980)$ and $I=1$ $a_0(980)$
productions in the same $D^0 \to \bar{K}^0 R$ decay.

When it comes to compare with the experiment we can see that the
$f_0(980)$ signal is quite narrow and it is easy to extract its
contribution to the branching ratios by assuming a smooth background
(shown in Fig.~\ref{Fig:dgamr600} by the dotted line) below the
$f_0(980)$ peak as a continuation of the $f_0(500)$ broad structure
at lower energies. For the case of the $\pi^0 \eta$ distribution we
get a clear peak that we associate to the $a_0(980)$ resonance,
remarkably similar in shape to the one found in the
experiment~\cite{Rubin:2004cq}. Yet it is obvious that not all the
strength seen in Fig.~\ref{Fig:dgamr600} can be attributed to the
$a_0(980)$ resonance. One should recall that the chiral unitary
approach provides amplitudes, in this case the $\pi^0 \eta$
amplitude, but the amplitudes provide poles that one associates to
resonances but also background contributions, and this is the case
of the $\pi^0 \eta$ distribution. In order to get a "$a_0(980)$"
contribution we subtract a smooth background that we depict by a
open dotted line in the figure. By doing that we have a remaining
"resonant" shape with an apparent width of $80$ MeV, which is in the
middle of the $50 - 100$ MeV of the PDG~\cite{pdg}. Integrating the
area below these structures we obtain
\begin{eqnarray}
R &=& \frac{\Gamma(D^0 \to \bar{K}^0 a_0(980), a_0(980) \to \pi^0
\eta)}{\Gamma(D^0 \to \bar{K}^0 f_0(980), f_0(980) \to \pi^+\pi^-)}
\nonumber \\
& = & 6.7 \pm 1.3 , \label{ratioth}
\end{eqnarray}
where we have added a $20\%$ theoretical error due to uncertainties
in the extraction of the background.

Experimentally we find from the PDG and the
Refs.~\cite{Muramatsu:2002jp,Rubin:2004cq},
\begin{eqnarray}
&& \Gamma(D^0 \to \bar{K}^0 a_0(980), a_0(980) \to \pi^0 \eta)
\nonumber \\
&&= (6.5 \pm 2.0) \times 10^{-3}, \\
&& \Gamma(D^0 \to \bar{K}^0 f_0(980), f_0(980) \to
\pi^+\pi^-)\nonumber \\
&&= (1.22^{+0.40}_{-0.24}) \times 10^{-3} .
\end{eqnarray}

The ratio that one obtains from there is
\begin{eqnarray}
R = 5.33^{+2.4}_{-1.9} . \label{ratioex}
\end{eqnarray}

The agreement found between Eq.~(\ref{ratioth}) and
Eq.~(\ref{ratioex}) is good, within errors. This is, hence, a
prediction that we can do parameter free.

\begin{figure}[htbp]
\begin{center}
\includegraphics[scale=0.44]{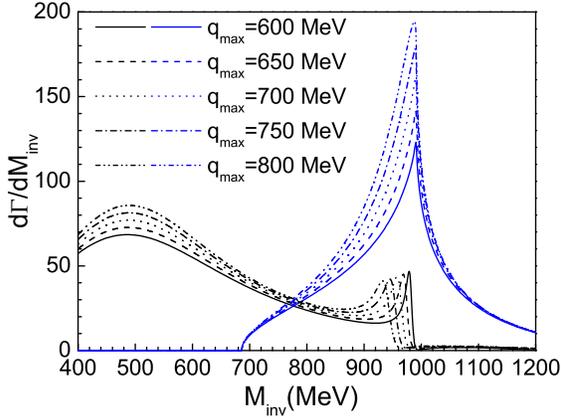}
\caption{(Color online) The $\pi^+ \pi^-$ (black curves) and $\pi^0
\eta$ (blue curves) invariant mass distributions with different cut
off $q_{\rm max}$ for the $D^0 \to \bar{K}^0 \pi^+ \pi^-$ decay and
$D^0 \to \bar{K}^0 \pi^0 \eta$ decay, respectively.}
\label{Fig:dgamr}
\end{center}
\end{figure}

As we mentioned, the explicit consideration of the $\eta \eta$
channel in the meson meson interaction, required to use a cut off
$q_{\rm max} = 600$ MeV~\cite{weihong} to agree with experimental
amplitudes, smaller than in Ref.~\cite{npa} where this channel was
omitted. We use the same cut off here. Yet, we want to show
explicitly that the ratio obtained does not get spoiled even if a
wide range of cut offs are used. In Fig.~\ref{Fig:dgamr}, we show
the results for five different, higher values of $q_{\rm max}$. The
magnitude of the $a_0(980)$ production grows a bit with $q_{\rm
max}$, with the prescription taken above, but the strength of the
$f_0(980)$ production also grows as a consequence of an increase in
the width. One can also see that the peak of the $f_0(980)$ moves to
lower energies, what puts constraints on $q_{\rm max}$, but we see
that, even within this broad range of values of $q_{\rm max}$, the
ratio of Eq.~(\ref{ratioth}) remains within the errors of this
equation and is a solid prediction.

It should not go unnoticed that we also predict a sizeable fraction
of the decay width into $D^0 \to \bar{K}^0 f_0(500)$, with a
strength several times bigger than for the $f_0(980)$. The $\pi^+
\pi^-$ distributions is qualitatively similar to that obtained in
Ref.~\cite{weihong} for the $\bar{B}^0 \to J/\psi \pi^+\pi^-$ decay,
although the strength of the $f_0(500)$ with respect to the
$f_0(980)$ is relatively bigger in this latter decay than in the
present case (almost $50\%$ bigger). The $\bar{B}^0 \to J/\psi
f_0(500),~f_0(500) \to \pi^+ \pi^-$ decay mode, together with the
$f_0(980)$ one have been identified in Ref.~\cite{Aaij:2014siy}
through a partial wave analysis, and the rates obtained are
comparable with the findings of Ref.~\cite{weihong}. Such a partial
wave analysis is not available from the work of
Ref.~\cite{Muramatsu:2002jp}, where the analysis was done assuming a
resonant state and a stable meson, including many contributions, but
not the $K^0_s f_0(500)$. Yet, a discussion is done at the end of
the paper~\cite{Muramatsu:2002jp} in which the background seen is
attributed to the $f_0(500)$. With this assumption they get a mass
and width of the $f_0(500)$ compatible with other experiments.
Further analyses in the line of~\cite{Aaij:2014siy} would be most
welcome to separate this important contributions to the $D^0 \to
K^0_s \pi^+ \pi^-$ decay.

\section{Further considerations}

\begin{figure}[htbp]
\begin{center}
\includegraphics[scale=0.8]{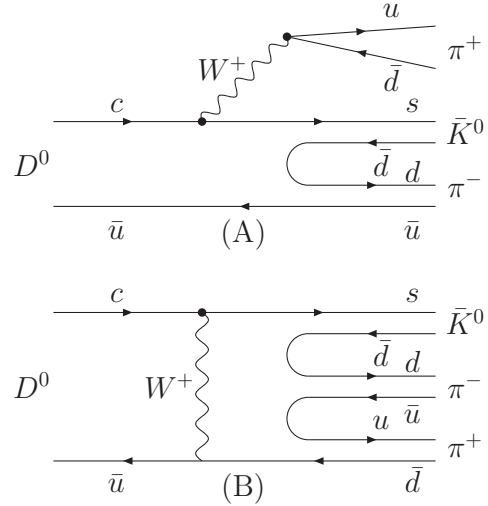}
\caption{External emission diagram [(A)] and the $W$-exchange
diagram [(B)] for $D^0 \to \bar{K}^0 \pi^+ \pi^-$ decay}
\label{wexchange}
\end{center}
\end{figure}

Our results are based on the dominance of the quark diagrams of
Fig.~\ref{Fig:feyn}. In the weak decay of mesons the diagrams are
classified in six different
topologies~\cite{Chau:1982da,Chau:1987tk}: external emission,
internal emission, $W$-exchange, $W$-annihilation, horizontal
$W$-loop and vertical $W$-loop. As shown in
Ref.~\cite{Cheng:2010vk}, only the internal emission graph
(Fig.~\ref{Fig:feyn} of the present work) and
$W$-exchange~\footnote{The $W$-exchange and $W$-annihilation are
often referred together as weak annihilation diagrams.} contribute
to the $D^0 \to \bar{K}^0 f_0(980)$ and $D^0 \to \bar{K}^0a_0(980)$
decays. In Ref.~\cite{kaminski} the $D^0 \to \bar{K}^0 \pi^+\pi^-$
decay is studied. Hence, only the $D^0 \to K^0_s f_0(980)$ decay can
be addressed, which is accounted for by proper form factors and
taken into account by means of the $M_2$ ($K^0_s[\pi^+\pi^-]_s$)
amplitude, which contains the tree level internal emission, and
$W$-exchange (also called annihilation mechanism). In order to
establish connection with the work of Ref.~\cite{kaminski}, let us
draw the external emission and $W$-exchange diagrams pertinent to
the $D^0 \to \bar{K}^0\pi^+\pi^-$ decay, as shown in
Fig.~\ref{wexchange}.

It is also instructive to recall the basic non-leptonic Hamiltonian
at the quark level responsible for this
transition~\cite{Buchalla:1995vs,ElBennich:2009da,Leitner:2010ai}
\begin{eqnarray}
H_{\rm W} &=& \frac{G_F}{\sqrt{2}} V_{cs} V_{ud} \bar{c}\gamma_{\mu}
(1   -   \gamma_5)s \bar{d}\gamma^{\mu} (1  - \gamma_5)u \nonumber
\\ &&  +   h.c.  \label{eq:hweak}
\end{eqnarray}
This Hamiltonian transforms as an isospin $I=1$ operator.
Consequently the decay amplitude of $D^0 \to K\pi\pi$ is
\begin{eqnarray}
T(D^0 \! \to \! K\pi\pi) \! =  <K^0_s M_1 M_2|H_{\rm W}|D^0>,
\end{eqnarray}
where the two meson system $M_1M_2$ ($\pi^+\pi^-$ here) can have
$I=0,~1,~2$. This is the case in the diagram of Fig.~\ref{wexchange}
(A) where the $c \bar{u}$, $\pi^+$ intermediate state can have
$I=1/2,~3/2$, which allows the $\pi^+\pi^-$ system to have
$I=0,~1,~2$ in the final $\bar{K}^0 \pi^+\pi^-$ state. However, the
diagram of Fig.~\ref{wexchange} (A) will not contribute to our
resonance production which requires the $\pi^+\pi^-$ $S$ wave loop,
as seen in Fig.~\ref{Fig:mesonmesonFSI}, due to the vector structure
of Eq.~(\ref{eq:hweak}) in the $csW^+$ vertex of
Fig.~\ref{wexchange} (A). This is also the case in the
phenomenological analysis of Ref.~\cite{Cheng:2010vk}. Then, in the
remaining mechanisms of Fig.~\ref{Fig:mesonmesonFSI} and
Fig.~\ref{wexchange} (B) the $\pi^+\pi^-$ can only be in $I=0$ or
$1$.

In our study we have isolated the $S$ wave of the pions in order to
get the $f_0(500)$, $f_0(980)$ resonances, and the $a_0(980)$ in the
case of $\pi^0 \eta$. Certainly, the operator of
Eq.~(\ref{eq:hweak}) allows other angular momenta, and indeed
experimentally $\rho$ meson and other mesons can be obtained, but
the experimental analysis of
Refs.~\cite{Muramatsu:2002jp,Rubin:2004cq} with partial wave
analysis separate the contributions of $f_0(980)$ and $a_0(980)$
production, which allows us to compare directly with these data
without the need to look into other channels. Also, although in
principle the amplitudes depend on two independent Mandelstam
variables as seen in Ref.~\cite{kaminski}, the fact that we do not
consider the $\bar{K}^0 \pi^-$ interaction (leading to the
$\kappa$), which would just provide a background in the $\pi^+\pi^-$
mass distribution for the reasons discussed at the end of section
II, makes our amplitude dependent upon the invariant mass of
$\pi^+\pi^-$ or $\pi^0 \eta$.

Concerning the $W$-exchange diagrams, which we have ignored in our
approach, we would like to argue in favor of its relative smallness
with two arguments: firstly, in Fig.~\ref{Fig:feyn} (A) we can see
that the $\bar{u}$ quark of the $D^0$ is a spectator. We thus have a
one body operator at the $D^0$ quark level. However, in the
$W$-exchange one involves the two quarks of $D^0$ and the amplitude
squared involves the probability to find two quarks, smaller than
that of finding one quark. This situation is typical in nuclear
reactions, where the $W$-exchange would have its equivalent in the
exchange currents~\cite{Gil:1997bm}. The second argument is that in
the $W$-exchange diagram of Fig.~\ref{wexchange} (B) there is a
double hadronization compared to the single hadronization of
Fig.~\ref{Fig:feyn} (B). The hadronization reverts into a decreased
rate for two meson production compared to the single meson of the
original $q\bar{q}$, which we can estimate in about one order of
magnitude from the experimental rate~\cite{pdg,LHCb:2012ae} (see
Ref.~\cite{weihong} for details),
\begin{eqnarray}
\frac{\Gamma (\bar{B}^0_s \to J/\psi f_0(980); f_0(980) \to
\pi^+\pi^-)}{\Gamma (\bar{B}^0_s \to J/\psi \phi)} = 0.14 . \nonumber \\
\end{eqnarray}

In the literature there is much discussions about the relevance of
the $W$-exchange mechanism. In Ref.~\cite{Cheng:2010vk} an empirical
analysis is done based on giving a weight to the different
topological mechanisms, and the $W$-exchange mechanism (evaluated
under the assumption that the $f_0$ and $a_0$ resonances are
$q\bar{q}$ or tetraquark states) appears of the same order of the
internal conversion, with opposite sign, that makes the $C-E$
combination in $a_0$ production bigger than the $C+E$ combination in
$f_0$ production.~\footnote{The $C$ and $E$ are the contributions of
the internal conversion and $W$-exchange, and $C-E$ and $C+E$ the
combinations found in Ref.~\cite{Cheng:2010vk} for $a_0$ and $f_0$
production, respectively.} However, in the same paper, a
factorization approach is followed (see section V of
Ref.~\cite{Cheng:2010vk}) in which the $W$-exchange contribution is
claimed to be suppressed and is neglected in that approach. The
present work neglects the $W$-exchange mechanism and produces a
large $a_0(980)$ production relative to $f_0(980)$ due to the
mechanism of final state interaction. We should note that in both
cases, the intermediate production of $K\bar{K}$ states, and further
rescattering to give $\pi^+\pi^-$ or $\pi^0 \eta$ in the final
states, is a novelty of our approach compared to other approaches
and an essential ingredient in the results due to the strong
coupling of the $f_0(980)$ and $a_0(980)$ resonances to $K\bar{K}$.

The dominance of the internal emissions in this kind of processes is
also supported in other
works~\cite{Stone:2013eaa,Aaij:2014siy,Li:2011pg,Aaltonen:2011nk,Abazov:2011hv}.
In Ref.~\cite{kaminski} a detailed discussion is made of results in
different works. The $W$-exchange mechanism in Ref.~\cite{kaminski}
depends on two unknown form factors which are fitted to the data and
a phase which is unknown. From a fit to the data, a minimal strength
of about $20\%$ is obtained for the $W$-exchange mechanism,
suggesting that the contribution could be bigger. It is clear that
this issue is still open but the relative smallness of the
$W$-exchange mechanism has many arguments in favor, and our study,
producing a big ratio of $a_0(980)$ versus $f_0(980)$ production due
to final state interaction in coupled channels, neglecting the
$W$-exchange mechanism, provides extra support for its smallness.
Note that this $a_0/f_0$ large ratio was the main reason of the
relatively large weight of the $W$-exchange mechanism in the fit of
Ref.~\cite{Cheng:2010vk}. Studies along the lines of
Ref.~\cite{kaminski} for $D^0 \to \bar{K}^0 \pi^0 \eta$ would help
bring extra light into this issue.

\section{Summary and conclusions}

We have studied the decay of the  $D^0$ decay into $K^0_s$ and a
scalar resonance, $f_0(500)$, $f_0(980)$, $a_0(980)$. For this
purpose we have identified the weak mechanism that allows the
formation of a $\bar K^0$, that will act as a spectator,  and a pair
of mesons, $K\bar{K}$, $\pi \pi$, $\pi^0 \eta$, $\eta \eta$, etc.,
that upon interaction will give rise to the $f_0(500)$, $f_0(980)$,
$a_0(980)$ resonances. The first step is the production of a
$\bar{K}^0_s$ and a pair of $q \bar q$, which upon hadronization
leads to these pairs of mesons. The hadronization is done in an easy
way, by looking at the flavor content in meson meson of the
hadronized $q \bar q$ pair. This is sufficient in the present case
where we only aim at determining the shape of the invariant mass
distributions and the relative weight of the different production
modes, but not absolute rates. Once the weight of the different
$\bar K^0$-meson-meson components has been determined we then allow
these meson-meson components to interact, using for it the chiral
unitary approach, and they give rise to the $f_0(500)$, $f_0(980)$,
$a_0(980)$ resonances. They are seen in the $\pi^+ \pi^-$ invariant
mass distributions [$f_0(500)$, $f_0(980)$] and the $\pi^0 \eta$
distribution [$a_0(980)$], and we not only get the poles of these
resonances but also realistic mass distributions that can be
compared with experiment. We found the shape of the $\pi^0 \eta$
distribution rather similar to the one found in the experiment, and
we obtained a ratio of the branching ratios for $a_0(980)$ and
$f_0(980)$ production in good agreement with experiment, all of it
accomplished without any free parameter, meaning that the parameters
of the theory have been determined before hand in the study of the
meson meson interaction.

We emphasized the fact that it is the nature of the weak
interactions, that allows for isospin violations, what made possible
the production of the $a_0(980)$ and $f_0(980)$ resonances in the
same decay. This is a most welcome feature that has allowed to test
simultaneously the production of the two resonances in the same
reaction offering new test for the chiral unitary approach than
allowed in strong interaction  reactions, providing yet one more
example of support for the dynamically generated nature of the low
lying scalar mesons.

\section*{Acknowledgments}
One of us, E. O., wishes to acknowledge support from the Chinese
Academy of Science (CAS) in the Program of Visiting Professorship
for Senior International Scientists. This work is partly supported
by the Spanish Ministerio de Economia y Competitividad and European
FEDER funds under the contract number FIS2011-28853-C02-01 and
FIS2011-28853-C02-02, and the Generalitat Valenciana in the program
Prometeo, 2009/090. We acknowledge the support of the European
Community-Research Infrastructure Integrating Activity Study of
Strongly Interacting Matter (acronym HadronPhysics3, Grant Agreement
n. 283286) under the Seventh Framework Programme of EU. This work is
also partly supported by the National Natural Science Foundation of
China under Grant Nos. 11105126, 11375080, and 10975068, and the
Natural Science Foundation of Liaoning Scientific Committee under
Grant No 2013020091. The Project Sponsored by the Scientific
Research Foundation for the Returned Overseas Chinese Scholars,
State Education Ministry.

\bibliographystyle{plain}

\end{document}